# Some more Exotic Dark Matter Candidates: GUT Balls, Fermi Balls…

*C Sivaram*

Indian Institute of Astrophysics, Bangalore, 560 034, India

Telephone: +91-80-2553 0672; Fax: +91-80-2553 4043

e-mail: sivaram@iiap.res.in

*Kenath Arun*

Christ Junior College, Bangalore, 560 029, India

Telephone: +91-80-4012 9292; Fax: +91-80- 4012 9222

e-mail: kenath.arun@cjc.christcollege.edu

**Abstract:** The nature of dark matter (DM), which is supposed to constitute about one-fourth of the universe, is still a mystery. There is evidence that much of the DM may be made up of as yet undiscovered particles with several experiments all over the world trying to detect these. In this article we introduce some new candidates which are exotic in nature but still are consistent with known physics. We look at DM objects that can be formed by the balance of gravity with the four-fermion force, nuclear tension, etc. We see that their radii are much larger than their corresponding Schwarzschild radius; hence they are distinct from Hawking primordial black holes. We have determined their mass and the required number densities to account for the DM in the galaxy and possible ways of detecting them.

## 1. Introduction

A continuing cosmological conundrum concerns the constitution and nature of Dark Matter currently believed to compose about a fourth of the total matter in the universe (Sivaram, 1987; 1985). Although there is also the problem of missing baryonic matter (about forty percent of baryons are believed missing from the nearby universe (Day, 2008; Danforth & Shull, 2008)) which perhaps mostly exists in the sparse intergalactic medium as indicated by recent UV surveys. (Werner et al, 2008)

It appears that matter in the form of MACHOS, compact stellar remnants, etc. account for only a fraction of the requisite amount of DM. Thus there is compelling evidence that much of the DM may be made up of as yet undiscovered particles like axions, neutralinos, gravitinos or composites of the same. (Sivaram & Arun, 2011)

Indeed several experiments in many underground laboratories all over the world have been trying to detect wimps (weakly interacting massive particles), axions, and even wisps (weakly interacting slim particles) (Hoogeveen & Ziegenhagen, 1991; Mueller et al., 2009; Redondo & Ringwald, 2011). Many of these ongoing experimental endeavours are very ingenious set ups.

Nevertheless there is so far no unambiguous evidence for any of these postulated particles. Astrophysical evidence from detection of gamma-rays (or other radiation) by decay of these massive DM particles forming compact objects or clumps (perhaps in the galactic centre) have also been claimed recently (Sivaram & Arun, 2011; Diemand, Moore & Stade, 2005).

Many of these conjectured particles are in the preferred range of 100 GeV to a TeV (Sivaram & Arun, 2011; Diemand, Moore & Stade, 2004). There could be DM objects or clumps made up of these particles bound by their mutual self gravity (Sivaram & Arun, 2011). Limits have already been placed on the abundance or density of these objects. (Sivaram & Arun, 2011; Diemand, Moore & Stade, 2004)

Considering that the nature of the DM is still entirely unknown and there could be other candidate objects, we are unapologetic in this article about introducing some NEW

CANDIDATES which are also exotic in nature but still are consistent with known physics and laws of physics as currently understood. In earlier papers we have invoked electroweak DM particles having a mass (Sivaram, 1999; 1995):

$$m_{EWDM} = \frac{e\hbar}{G_F^{1/2} c} = 5 \times 10^{-23} g \approx 30 GeV \qquad \ldots (1)$$

(*e* being the electron electric charge, $G_F$ the universal Fermi weak interaction constant $\approx 1.4 \times 10^{-49} erg cm^3$, $\hbar$ and *c* are as usual Planck's constant and vacuum light velocity)

(We may mention in passing that the claimed DAMA detection and gamma rays from DM decays in the galaxy (as claimed) imply masses of order ~60 GeV perhaps a pair of particles with above mass annihilating! (Gelmini, 2006; Sivaram, 1994a))

As the mass given in eq. (1) is made up of only basic constants it could be significant. Again in references (Sivaram, 1999; 1995), we had also conjectured that the production of DM particles of the above mass in the early universe could at the electroweak TeV phase transition, generated vacuum energy which at the present epoch, would constitute a vacuum dark energy density (DE) given by:

$$\rho_{DE(present)} = \frac{e^7 \hbar^{17/2}}{G_F^{7/2}} \frac{G^{3/2}}{c^{1/2}} \approx 10^{-29} g cm^{-3} \qquad \ldots (2)$$

which corresponds to a cosmological constant $\Lambda = 10^{-56} cm^{-2}$ and a Hubble constant of $70 km/s/Mpc$ (Sivaram, 1994a; 1986), precisely as observed!

In terms of $m_{DM}$ particle as given in eq. (1), this implies:

$$\rho_{DE(present)} \approx \frac{m_{DM}^7 c^6}{\hbar^6} \left(\frac{\hbar G}{c^3}\right)^{3/2} \qquad \ldots (3)$$

Note the $m^7$ dependence! This shows that this is a more or less unique value which gives constancy with observations. This unification of DM and DE has been further explored in references (Sivaram, Arun & Nagaraja, 2011; Sivaram, 1994a).

## 2. Fermi Balls

Based again on the balance between weak interactions and gravity we propose here, so called '*Fermi balls*' arising from long-range two-neutrino exchange between two electrons or protons. This force has a $1/r^5$ dependence on the potential and has been involved in various contexts thus having a long tradition (Sivaram, 1983; Ivanenko & Tamm, 1934; Hartle, 1972). For $N$ particles in a spherical configuration of radius $R$ the force is of the form (it arises purely from the Fermi-four-fermion of a neutrino pair exchanged between massive fermions with a weak-coupling strength (Sivaram, 1983; Hartle, 1972)

$$F_W \approx \frac{G_F^2 N}{4\pi^3 R^5 \hbar c} \qquad \ldots (4)$$

This is to be balanced by the gravitational self energy scaling as $N^2$ with $1/R$ dependence, i.e. ($m$ being the particle mass, thus $Nm$ is the total mass)

$$F_G \approx \frac{GN^2 m^2}{R} \qquad \ldots (5)$$

Indeed eq. (4) after evaluation of the constants gives:

$$F_W \approx \frac{3 \times 10^{-83} N}{R^5}$$

Balancing the above forces, i.e. four-fermion and gravity force (overall neutrality would imply no Coulomb forces), gives a unique mass-radius ($M$-$R$) relation for those objects as:

$$MR^4 = \frac{G_F^2}{\hbar c G m} \qquad \ldots (6)$$

(Note the similarity to white-dwarf, $M$-$R$ relation $MR^3 = \frac{\hbar^6}{G m_P^{5/3} m_e}$ or for a neutron star supported by neutron degeneracy pressure $MR^3 = \frac{\hbar^6}{G m_n^{8/3}}$).

Thus for a configuration of mass $M$, the radius is given by:

$$R^4 = \frac{G_F^2}{\hbar c G m M} \qquad \ldots (7)$$

For *m ~1GeV* a configuration of 1 fermi radius (i.e. $R \sim 10^{-13}$ cm) would have a mass of a kilogram. An object of *0.1 fm* radius would have mass $\sim 10^4$ Kg, as $M \propto R^{-4}$

We call such objects FERMI BALLS.

To account for the total DM in our galaxy we need to invoke about $10^{40}$ Fermi Balls, having a galactic number density of $10^{-20} m^{-3}$. However, we note that the radius $R$ of these objects is much larger than their corresponding Schwarzschild radius, i.e. $R \gg R_{Sch}$

Thus these objects are NOT Hawking primordial black holes (PBH) although their radius of 1 fermi is comparable to Hawking PBH's! These objects became black holes only for $R < 10^{-16}$ cm. The detection methods for such objects are open and we would welcome suggestions. We hope to explore this in a later work.

In the case of FERMI BALLS above, we had the interplay of gravity and the long-range two neutrino four-fermion interaction which (from phase space) had a $1/r^5$ potential dependence. Can we have objects bound by nuclear force and gravity? The balance between surface tension and gravity plays a seminal role in many areas of physics and its applications [even in biophysics (motion of insects, capillary action in trees even limiting their height, Marangoni convection, etc.)].

Some time back, there was a discussion of why many small asteroids, (treated as icy-balls) have a radius of around fifty metres. This could be explained in terms of objects bound by surface tension and gravity.

If $S$ be the surface tension and $R$ be the radius, $\rho$ the density, then the surface energy is $S \cdot 4\pi R^2$ and the gravitational energy is $G \cdot \frac{16\pi^2}{9} \cdot \rho^2 R^5$ and balance between these forces (no viscosity or charge) gives for the radius (at which they are comparable)

$$R^3 \sim \frac{S}{G\rho^2} \qquad \ldots (8)$$

(*S* for water –ice $\sim 10^2$ dyne/cm, $\rho \sim 1$ gn cm$^{-3}$ which gives *R ~50 metres*)

This also gives an asteroid mass $\sim 10^6 - 10^7$ tons, implying they could cause much damage!

## 3. Nuclear Balls

Now, as is well known, nuclear forces, as for example in the liquid-drop model behave like a fluid with a surface tension of:

$$S_{Nucl} \approx 10^{21} dyne/cm \qquad \ldots (9)$$

This is a typical nuclear surface force increasing with area $a$ as $R^2$ (and with mass number as $A^{2/3}$). The density of the nuclear fluid is:

$$\rho_{Nucl} \approx 10^{13} gcm^{-3} \qquad \ldots (10)$$

Thus we could have '*Nuclear Balls*'.

For the above values of $S$ and $\rho$ as given by eq. (9) and (10), the radii of these balls are given as:

$$R_{NB} \approx \left(\frac{S_{Nucl}}{G\rho_{Nucl}^2}\right)^{1/3} \sim 1 metre \qquad \ldots (11)$$

This would give them a mass $\approx 4 \times 10^{11} - 10^{12} tons$. The radius of these nuclear chunks is again $>>R_{Sch}$, so they are not black holes.

We would require about $\sim 10^{27}$ of these 'nuclear balls' to account for the galactic DM. This implies about one such object in a volume of our solar system. We also note that their mass corresponds to that of an asteroid but their radius is far smaller. It would be interesting to look for such an object in the asteroid belt!

## 4. EW Balls and GUT Balls

These 'nuclear balls' could be formed in the early universe when densities were comparable to nuclear densities (in a sense they are similar to 'quark nuggets' which have also been postulated as DM candidates.

However, at the earlier epoch of the electroweak transition, we could have the formation of similar objects. Now the corresponding density would be that corresponding to the EW scale $\sim 10^2$ *GeV*, i.e. $\rho_{EW} \approx 10^{26} gcm^{-3}$. The corresponding 'tension', analogous to $S_{Nucl}$ in eq. (9) above is (Sivaram, 1994b; 1986) $T_{EWs} \approx 10^{32} dyne/cm$.

The corresponding radius of this '*electroweak gravity ball*' or EWB is:

$$R_{EWB} \approx \left(\frac{T_{EWB}}{G\rho_{EW}^2}\right)^{1/3} \sim 10^{-4} \, cm \qquad \ldots (12)$$

So these gravitating EW balls have about micron radius and weigh about $4 \times 10^{14}$ gm. So as before, their radius is much larger than their Schwarzschild radius and they are thus not Hawking black holes although their mass is intriguingly close to the Hawking PBH (whose evaporation time is comparable to the Hubble age), but their radius much larger.

We need about $10^{30}$ of these EW balls for our galactic DM which again implies one such object in a solar system volume.

We could also have such objects forming the GUTS phase transition. In that case, they would be '*GUT balls*', with a tension $T_{GUT} \approx 10^{70} \, dyne/cm$ (Sivaram, 1990) implying a much smaller radius $R \sim 10^{-20}$ cm. $T_{GUT}$ would depend on the GUT scale, i.e. $T \propto M_{GUT}^{-3}$ (Sivaram, 1990; 1994b)

In short, there is a plethora of possibilities for new varieties of exotic DM objects, all consistent with basic physical laws, underlying fundamental interactions in nature.